\documentclass[12pt,letterpaper]{article}

\usepackage{amsmath, amssymb, amsfonts, amsthm}
\usepackage{graphicx}
\usepackage{hyperref}
\usepackage{natbib}
\usepackage{booktabs, multirow}
\usepackage{listings}
\usepackage{adjustbox}
\usepackage{float}
\usepackage{enumitem}
\usepackage{algorithm, algorithmic}

\usepackage{setspace}

\usepackage{geometry}
\theoremstyle{definition}

\newtheorem{theorem}{Theorem}

\hypersetup{
    colorlinks=true,
    linkcolor=blue,
    filecolor=magenta,
    urlcolor=cyan,
    citecolor=blue
}
\title{AdaptICA: Data-Adaptive Transformation Learning for Independent Component Analysis}
\author{Lida Jalili$^{1}$, Jingyu Liu$^{2}$, Vince D. Calhoun$^{3}$, Li-Hsiang Lin$^{1*}$ \\
$^{1}$Department of Mathematics and Statistics, Georgia State University,\\ Atlanta, Georgia 30303, United States\\ $^{2}$Department of Computer Science, Georgia State University,\\ Atlanta, Georgia 30303, United States\\ $^{3}$Tri-institutional Center for Translational Research in Neuroimaging and\\ Data Science, Georgia State University, Georgia Institute of Technology,\\ Emory University, Atlanta, Georgia 30303, United States \\
$^{*}$Corresponding Author: $lhlin@gsu.edu$}
\date{\today}

\begin{document}

\maketitle
\vspace{-0.5 cm}
\begin{abstract}
Independent component analysis (ICA) is widely used to recover latent structure from signal and imaging data, but standard ICA assumes that the observed measurement scale preserves a linear mixing structure. This assumption may fail for features produced through nonlinear preprocessing, such as band-specific power in motor-imagery EEG. We propose AdaptICA, an adaptive transformation-based framework that jointly learns grouped componentwise transformations and the demixing structure using a profiled mutual-information criterion. Because the transformation and demixing parameters may compensate for one another, their joint estimation introduces new identifiability and asymptotic challenges. We establish identifiability, consistency, and asymptotic normality of the transformation estimator, together with joint strong consistency of the transformation and demixing estimators. AdaptICA selects the transformation structure data-adaptively and includes the identity transformation as a candidate, thereby reducing to standard ICA when no scale adjustment is needed. Extensive simulations support the theoretical results. Applications demonstrate that AdaptICA can recover more independent and interpretable sources when transformation is beneficial while retaining standard ICA when the original measurement scale is adequate.
\end{abstract}
key words: Biomedical signal processing, Blind source separation, Neuroimaging, Box-Cox transformation, Nonlinear distortion
\thispagestyle{empty}
  
\setstretch{2}
\newpage
\setcounter{page}{1}
\section{Introduction}
\label{sec:intro}

Independent component analysis (ICA) is a fundamental method for unsupervised
representation learning and blind source separation in machine learning,
statistics, signal processing, and data science
\citep{comon1994independent,cardoso1999high,hyvarinen2000independent,
lee2003specialissue,cardoso2003dependence}. By representing multivariate
observations through statistically independent latent components, ICA and its
extensions support feature extraction, dimension reduction, denoising,
latent-structure discovery, and downstream learning
\citep{hyvarinen1999fast,cardoso1993blind,bell1995information,
hyvarinen2001independent,bach2003beyond,learnedmiller2003ica,
matteson2017dCovICA,sarela2003overlearning,sarela2005denoising,
pfister2019robustifying,sasaki2022representation}. Its identifiability,
estimation, and asymptotic properties have also been developed through
M-estimation and semiparametric theory
\citep{chen2006efficient,chen2008order,ilmonen2011semiparametric,
virta2016jade}. These capabilities are particularly relevant to neuroimaging,
where measurements from many sensors or spatial locations contain overlapping
contributions from latent neural activity, physiological artifacts, and
measurement noise. Accordingly, ICA is routinely used to remove artifacts,
identify functional networks, and construct interpretable representations from
EEG, MEG, and fMRI data
\citep{makeig1996eeg,vigario2000independent,calhoun2001method,
beckmann2004probabilistic,calhoun2009review}, with related applications in
other high-dimensional imaging problems \citep{nascimento2005does}. Given this
broad use, it is important to determine whether the analyzed data
representation preserves the linear mixing structure on which ICA depends.

This requirement can be difficult to verify because ICA is often applied to
derived features rather than directly to the original sensor measurements,
and nonlinear feature construction may alter the underlying mixture structure.
Our motivating example is motor-imagery EEG, where physiologically meaningful
mu-band power is obtained by filtering, squaring, and averaging sensor-level
voltages \citep{pfurtscheller1999event}. Because these nonlinear operations are
applied after the latent neural signals have been mixed at the scalp sensors,
the resulting power variables need not satisfy the linear representation
assumed by standard ICA. The observed dependence structure may be further
affected by physical mechanisms such as volume conduction and field spread
\citep{nunez2006electric,hamalainen1993meg,ilmoniemi2019brain}. A common
practice is to apply a fixed logarithmic, square-root, or other transformation
before ICA, typically using the same transformation for all variables and
selecting it independently of the source-separation objective. Such
preprocessing does not determine whether the selected transformation improves
the independence of the recovered latent components.

Learning transformations within ICA creates additional statistical and
methodological challenges. The transformation and demixing parameters may
compensate for one another, so identifiability of the standard ICA model does
not automatically extend to the enlarged model. Moreover, because both sets of
parameters are estimated from the same data, consistency of the transformation
estimator alone does not guarantee consistency of the recovered demixing
structure, and the usual asymptotic theory for ICA cannot be applied without
further justification. Classical transformation methods primarily target
marginal normality, variance stabilization, probability coherence, or
regression additivity
\citep{box1964analysis,dobson2018introduction,lin2020transformation}, rather
than independence of latent components. General nonlinear and post-nonlinear
ICA models provide greater flexibility
\citep{taleb1999postnonlinear,jutten2003blind,harmeling2003postnonlinear}, but
they are typically non-identifiable without additional structural, temporal,
or auxiliary information
\citep{kagan1973characterization,sasaki2022representation}. A systematic
framework is therefore needed to learn transformations according to the ICA
objective while retaining identifiability, tractable estimation, joint
consistency, and valid asymptotic inference.

Motivated by this phenomenon observed in the motor-imagery EEG application, we
propose adaptive transformation-based independent component analysis
(AdaptICA), a framework that learns an appropriate measurement scale through
grouped componentwise transformations selected by a profiled
mutual-information criterion. Grouping allows related variables to share
information while accommodating heterogeneous distortions across groups. The
identity transformation is included in the model class, so standard ICA is
retained when no scale adjustment is supported by the data. We establish
identifiability of the transformation parameter, consistency and
\(\sqrt{n}\)-asymptotic normality of its estimator, an oracle property for the
demixing estimator, and joint almost-sure consistency of the transformation
and demixing estimators. Thus, AdaptICA provides a unified extension of linear
ICA that uses the independence objective itself to determine whether and how
the measurement scale should be transformed, without sacrificing the
interpretability and computational structure of standard ICA.

Beyond neuroimaging, the proposed framework is applicable whenever ICA is used
as an intermediate representation for denoising, feature extraction,
clustering, visualization, prediction, or latent-pattern discovery. By
combining transformation learning with an independence-based objective,
AdaptICA broadens the use of ICA to multivariate data whose analyzed scale is
affected by nonlinear preprocessing, feature construction, or unknown
componentwise distortions, while preserving standard ICA for datasets that
remain adequately linear on their original measurement scale.

The remainder of the paper is organized as follows.
Section~\ref{sec:method} introduces the AdaptICA model and estimation procedure.
Section~\ref{sec:theory} presents the identifiability and asymptotic results.
Section~\ref{sec:simulation-studies} reports the simulation studies, and
Section~\ref{sec:real_data} presents the EEG and MEG applications. Concluding
remarks and directions for future research are provided in the final section,
with proofs, additional simulations, and detailed real-data analyses included
in the Supplemental Material.

\section{Methodology}\label{sec:method}
Let \(\mathbf{X}\in\mathbb{R}^{n\times T}\) denote the observed data matrix,
where the \(i\)-th row \(\mathbf{x}_i\in\mathbb{R}^T\) represents one
observation measured over \(T\) variables, such as sensor channels. In classical ICA, each observation is assumed to follow the linear
mixing model
\[
\mathbf{x}_i=\mathbf{A}\mathbf{s}_i,\qquad i=1,\ldots,n,
\]
where \(\mathbf{A}\in\mathbb{R}^{T\times T}\) is an invertible mixing matrix
and \(\mathbf{s}_i\in\mathbb{R}^T\) is a latent source vector with mutually
independent components. Under standard ICA assumptions, the latent sources are
identifiable up to permutation and scaling. In many applications, however, the
observed variables are additionally subject to unknown componentwise nonlinear
distortions, so direct application of the ICA may fail to recover the
underlying independent structure. Our goal is therefore to recover the latent
independent components while accommodating such nonlinear effects through a
structured parametric transformation.

To address this issue, we introduce an invertible componentwise transformation
\(T_{\boldsymbol{\lambda}}:\mathbb{R}^T\to\mathbb{R}^T\) applied to each
observed vector. At the population level, we assume that the observed random
vector \(\mathbf{x}\) follows the \emph{transformation-based ICA model}
\begin{equation}
\mathbf{x}
=
T_{\boldsymbol{\lambda}_0}^{-1}(\mathbf{A}_0\mathbf{s}),
\label{eq:tica_model}
\end{equation}
where \(\boldsymbol{\lambda}_0\in\Lambda\subset\mathbb{R}^K\) is the true
transformation parameter, \(\mathbf{A}_0\in\mathbb{R}^{T\times T}\) is an
invertible mixing matrix, and
\(\mathbf{s}=(s_1,\ldots,s_T)^\top\) has mutually independent components.
Equivalently, applying the true transformation yields the linear ICA
representation
\(T_{\boldsymbol{\lambda}_0}(\mathbf{x})=\mathbf{A}_0\mathbf{s}\).
Thus, \(T_{\boldsymbol{\lambda}}\) removes componentwise nonlinear distortion
so that ICA can be performed on a measurement scale that better preserves a
linearly separable latent structure. Although this formulation accommodates a
general class of invertible transformations, for illustration and theoretical
development we adopt grouped parametric transformations based on the Box--Cox
family \citep{box1964analysis}, which provides a parsimonious and interpretable
specification while allowing related variables to share transformation
parameters. This parametric choice is not essential to the general AdaptICA
principle; more flexible nonparametric specifications, such as constrained
monotone spline transformations \citep{ramsay1988monotone}, can be incorporated.

For the observed data matrix \(\mathbf{X}\), let \(x_{ij}\) denote the \(j\)-th component of
\(\mathbf{x}_i\), for \(i=1,\ldots,n\) and \(j=1,\ldots,T\). Rather than
assigning a separate transformation parameter to each variable, we allow
variables to share parameters through a grouped structure. Specifically, we
partition the \(T\) variables into \(K\) groups \(G_1,\ldots,G_K\), with
typically \(K\le T\), and let \(G(j)\in\{1,\ldots,K\}\) denote the group index
of variable \(j\). We then define the transformation parameter vector
\(\boldsymbol{\lambda}=(\lambda_1,\ldots,\lambda_K)^\top\in\Lambda\), where
\(\Lambda\) is a compact parameter space, for example
\(\Lambda=[-2,2]^K\) for numerical stability. This grouped parameterization
reduces the transformation dimension from \(T\) to \(K\), thereby improving
parsimony and estimation stability. It is particularly natural in
neuroimaging applications, where observed variables such as EEG channels may be
organized into anatomically or functionally related regions and are therefore
expected to share similar nonlinear distortion patterns. As a concrete parametric family, we adopt the grouped Box--Cox transformation for demonstration.
For each observation \(\mathbf{x}_i\), define
\(\mathbf{z}_i=T_{\boldsymbol{\lambda}}(\mathbf{x}_i)\), whose \(j\)-th
component is
\begin{equation}
z_{ij}=T_{\lambda_{G(j)}}(x_{ij})=
\begin{cases}
(x_{ij}^{\lambda_{G(j)}}-1)/\lambda_{G(j)}, & \lambda_{G(j)}\neq 0,\\[4pt]
\log x_{ij}, & \lambda_{G(j)}=0,
\end{cases}
\qquad i=1,\ldots,n,\; j=1,\ldots,T.
\label{eq:boxcox_transform}
\end{equation}
The corresponding inverse transformation is
\begin{equation}
T_{\lambda}^{-1}(z)
=
\begin{cases}
(\lambda z + 1)^{1/\lambda}, & \lambda \neq 0,\\[4pt]
\exp(z), & \lambda = 0.
\end{cases}
\label{eq:boxcox_inverse}
\end{equation}
The parameter dimension $K$ carries significant practical implications for the model's interpretability and statistical performance. Rather than assuming independent transformations for every variable, we leverage the underlying structural properties of the data to group related sensors. For instance, in our Section~\ref{sec:real_data_BCI} analysis of EEG medical imaging data, the $T = 22$ individual channels are concisely represented by $K = 10$ parameters corresponding to distinct anatomical regions. By adopting this grouped configuration, the model achieves a more robust estimation of latent brain activity. This approach effectively exploits the spatial dependencies inherent in sensor locations, reducing the risk of overfitting while ensuring that the learned transformations align with known neurophysiological boundaries.


The fundamental premise is that, for an appropriate choice of $\boldsymbol{\lambda}$, each transformed observation vector $\mathbf{z}_i$ admits an approximately linear ICA representation. Given the transformed observations $\mathbf{z}_i = T_{\boldsymbol{\lambda}}(\mathbf{x}_i)$, we consider a linear demixing matrix $\mathbf{W} \in \mathbb{R}^{T \times T}$ and define the candidate components:
\begin{equation}
\mathbf{u}_i = \mathbf{W}\mathbf{z}_i = \mathbf{W} T_{\boldsymbol{\lambda}}(\mathbf{x}_i), \qquad i = 1,\ldots,n,
\label{eq:demixing}
\end{equation}
where $\mathbf{u}_i=(u_{i1},\ldots,u_{iT})^\top$ approximates the latent sources $\mathbf{s}_i$.  We quantify independence through mutual information (MI). Let
$f_{\mathbf{u}}$ denote the joint probability density function (pdf) of the
random vector $\mathbf{u}$, and let $f_{u_j}$ denote the marginal pdf of its
$j$-th component. The mutual information is
\begin{equation}
\mathrm{MI}(\mathbf{u})
=
\int_{\mathbb{R}^T}
f_{\mathbf{u}}(\mathbf{u})
\log\!\left\{
\frac{f_{\mathbf{u}}(\mathbf{u})}{\prod_{j=1}^T f_{u_j}(u_j)}
\right\}
\,d\mathbf{u}
=
D_{\mathrm{KL}}
\!\left(
f_{\mathbf{u}}
\;\Big\|\;
\prod_{j=1}^T f_{u_j}
\right),
\label{eq:MI_def}
\end{equation}
where $D_{\mathrm{KL}}(\cdot\|\cdot)$ denotes the Kullback--Leibler divergence.
Since $\mathrm{MI}(\mathbf{u})=0$ if and only if the components of
$\mathbf{u}$ are mutually independent, minimizing mutual information provides a
natural criterion for recovering independent sources.


At the population level, for transformed observations
$\mathbf{u}_i=\mathbf{W}T_{\boldsymbol{\lambda}}(\mathbf{x}_i)$, we may view
the target criterion as
$\mathrm{MI}\!\left(
\mathbf{W}T_{\boldsymbol{\lambda}}(\mathbf{x})
\right),$
that is, the mutual information of the random vector obtained after applying
the componentwise transformation and the linear demixing matrix. Joint
optimization over $(\mathbf{W},\boldsymbol{\lambda})$ is challenging because
the problem is nonconvex and because the transformation and demixing parameters
interact in a highly nonlinear way. We therefore adopt a profiled two-step
strategy.

Specifically, for any fixed $\boldsymbol{\lambda}\in\Lambda$, define the profiled population
objective
\begin{equation}
g(\boldsymbol{\lambda})
=
\inf_{\mathbf{W}}
\mathrm{MI}\!\left(
\mathbf{W}T_{\boldsymbol{\lambda}}(\mathbf{x})
\right),
\label{eq:profile_objective}
\end{equation}
which represents the smallest population dependence that can be achieved after
optimally demixing the transformed data. Under the true transformation
$\boldsymbol{\lambda}_0$, the transformed vector
$T_{\boldsymbol{\lambda}_0}(\mathbf{x})$ follows a linear ICA model, so the
population minimum is attained at zero up to the usual ICA indeterminacies.
Hence $g(\boldsymbol{\lambda})$ serves as the criterion for selecting the
transformation parameter. In practice, however, the population mutual information in
\eqref{eq:profile_objective} is unknown because it depends on the joint and
marginal densities of the latent demixed vector. We therefore replace
$g(\boldsymbol{\lambda})$ by a computable sample criterion
$\widehat g_n(\boldsymbol{\lambda})$, which we refer to as an empirical proxy
for the profiled mutual information. Concretely, for each candidate
$\boldsymbol{\lambda}$, we first transform the observed sample to obtain
$\mathbf{z}_i=T_{\boldsymbol{\lambda}}(\mathbf{x}_i)$, apply an ICA routine to
the transformed data to obtain a demixing estimate
$\widehat{\mathbf{W}}(\boldsymbol{\lambda})$, and then evaluate a sample
measure of residual dependence among the recovered components
$\widehat{\mathbf{u}}_i(\boldsymbol{\lambda})
=
\widehat{\mathbf{W}}(\boldsymbol{\lambda})
T_{\boldsymbol{\lambda}}(\mathbf{x}_i), i=1,\ldots,n.$
Thus,
$
\widehat g_n(\boldsymbol{\lambda})
=
\widehat{\mathrm{Dep}}_n
\!\left(
\widehat{\mathbf{u}}_1(\boldsymbol{\lambda}),
\ldots,
\widehat{\mathbf{u}}_n(\boldsymbol{\lambda})
\right),
$
where $\widehat{\mathrm{Dep}}_n$ denotes a sample dependence measure chosen to
approximate mutual information or, more generally, to rank candidate
transformations according to how close the recovered components are to
independence. For example, $\widehat{\mathrm{Dep}}_n$ may be implemented using
the empirical ICA contrast returned by FastICA, a nonparametric mutual
information estimator, or another dependence criterion such as distance
covariance. The essential requirement for the theory is that
$\widehat g_n(\boldsymbol{\lambda})$ converges uniformly to
$g(\boldsymbol{\lambda})$ over $\Lambda$.

We then estimate the transformation parameter by minimizing this sample
criterion:
\begin{equation}
\widehat{\boldsymbol{\lambda}}
=
\arg\min_{\boldsymbol{\lambda}\in\Lambda}
\widehat g_n(\boldsymbol{\lambda}).
\label{eq:lambda_estimator}
\end{equation}
In this way, Stage~1 selects the transformation that makes the data as close as
possible to admitting a linearly separable ICA representation. Given $\hat{\boldsymbol{\lambda}}$ in \eqref{eq:lambda_estimator}, Stage 2 extracts the final sources:
\begin{equation}
\mathbf{u}_i^\ast = \mathbf{W}^\ast T_{\hat{\boldsymbol{\lambda}}}(\mathbf{x}_i), \qquad i = 1,\ldots,n.
\label{eq:final_sources}
\end{equation}
We use FastICA in the final stage for efficiency and stability. The number of
groups \(K\) is chosen by the same profiled dependence criterion that defines
the transformation estimator in \eqref{eq:lambda_estimator}. Each candidate
value of \(K\) specifies a grouped parameterization
\(\boldsymbol{\lambda}\in\mathbb{R}^{K}\), and for that grouping we obtain
\(\widehat{\boldsymbol{\lambda}}(K)=\arg\min_{\boldsymbol{\lambda}\in\Lambda}
\widehat g_n(\boldsymbol{\lambda})\) with the associated criterion value
\(\widehat g_n(\widehat{\boldsymbol{\lambda}}(K))\). We consider a nested family
of anatomically defined groupings, ranging from a single common transformation
(\(K=1\)) through region- and hemisphere-based partitions to a separate
transformation for each channel (\(K=T\)), and select
\[
\widehat K
=\arg\min_{K}\ \widehat g_n\!\left(\widehat{\boldsymbol{\lambda}}(K)\right).
\]
Because \(\widehat g_n\) is the same profiled mutual-information proxy used in
\eqref{eq:lambda_estimator}, selecting \(K\) requires no additional criterion.
As an independent check that is not used in estimation or selection, we also
report the distance correlation among the recovered components at the selected
grouping; its reduction relative to standard ICA confirms that the improvement
is not an artifact of the mutual-information objective. The overall procedure
is summarized in Algorithm~\ref{alg:AdaptICA}.

\begin{algorithm}[H]
\caption{Transformation-based Independent Component Analysis}
\label{alg:AdaptICA}
\begin{algorithmic}[1]
\STATE \textbf{Input:} Data matrix $\mathbf{X}\in\mathbb{R}^{n\times T}$; search
set $\Lambda$; transformation $T_{\boldsymbol{\lambda}}$; dependence criterion
$\widehat g_n$; Stage 1 unmixing routine $\mathcal{A}$; candidate group counts
$\mathcal{K}$ with a grouping map for each $K\in\mathcal{K}$.
\STATE \textbf{Stage 1 (Profiled transformation and group selection):}
\FOR{each candidate $K \in \mathcal{K}$}
  \STATE Form the grouped parameterization $\boldsymbol{\lambda}\in\mathbb{R}^{K}$.
  \STATE Compute $\widehat{\boldsymbol{\lambda}}(K)=
         \arg\min_{\boldsymbol{\lambda}\in\Lambda}\widehat g_n(\boldsymbol{\lambda})$
         by transforming the data $\mathbf{z}_i=T_{\boldsymbol{\lambda}}(\mathbf{x}_i)$,
         applying $\mathcal{A}$, and evaluating
         $\widehat g_n(\boldsymbol{\lambda})=
         \widehat{\mathrm{Dep}}_n(\widehat{\mathbf U}(\boldsymbol{\lambda}))$.
\ENDFOR
\STATE Select $\widehat K=\arg\min_{K\in\mathcal{K}}
       \widehat g_n(\widehat{\boldsymbol{\lambda}}(K))$ and set
       $\hat{\boldsymbol{\lambda}}=\widehat{\boldsymbol{\lambda}}(\widehat K)$.
\STATE \textbf{Stage 2 (Final demixing and sources):}
Form $\mathbf{Z}^*\in\mathbb{R}^{n\times T}$ using $\hat{\boldsymbol{\lambda}}$.
Apply FastICA to $\mathbf{Z}^*$ to obtain $\mathbf{W}^*$ and source matrix
$\mathbf{U}^*\in\mathbb{R}^{n\times T}$ as in \eqref{eq:final_sources}; report
$\widehat{\mathbf{A}}=(\mathbf{W}^*)^{-1}$.
\STATE \textbf{Output:} $\hat{\boldsymbol{\lambda}}$, source matrix
$\mathbf{U}^*$, and $\widehat{\mathbf{A}}$.
\end{algorithmic}
\end{algorithm}
Algorithm~\ref{alg:AdaptICA} allows different unmixing routines in Stage~1, and our simulation studies examine several Stage~1 alternatives to evaluate their impact on transformation learning. In practice, we suggest using FastICA in both Stage~1 and Stage~2; under this choice, Stage~2 is in principle computationally redundant because Stage~1 already provides a demixing estimate at the selected $\hat{\boldsymbol{\lambda}}$. We nevertheless retain the explicit two-stage formulation for two reasons. First, it underscores the modular design of the method: Stage~1 is used only to select the transformation via a profiled dependence criterion and is not tied to any particular ICA solver, whereas Stage~2 performs final source extraction with a preferred algorithm. Second, this separation makes the profile structure transparent, which is essential for the identifiability, consistency, and oracle-property arguments developed in the theoretical section.




\section{Theoretical results}
\label{sec:theory}

Section~\ref{sec:method} introduces the profiled mutual-information criterion
for selecting the transformation parameter and the subsequent ICA step for final
source recovery. The main theoretical questions are whether the transformation
parameter is identifiable from the population criterion, whether the empirical
Stage~1 minimizer is consistent and efficient, and whether the final demixing
step behaves asymptotically as if the true transformation were known. This
section addresses these questions and provides the justification for the
two-step procedure in Algorithm~\ref{alg:AdaptICA}.

We assume that the data follow the transformation-based ICA model described in
Section~\ref{sec:method}, with true transformation parameter
\(\boldsymbol{\lambda}_0\), invertible mixing matrix \(\mathbf{A}_0\), and
latent source vector \(\mathbf{s}\) whose components are mutually independent,
with at most one Gaussian component. Throughout, the Box--Cox transformation,
its inverse, the transformed variables, the profiled population objective
\(g(\boldsymbol{\lambda})\), and the estimator
\(\widehat{\boldsymbol{\lambda}}_n\) are defined as in
Section~\ref{sec:method}. The final demixing matrix is obtained by applying ICA
to the sample transformed at \(\widehat{\boldsymbol{\lambda}}_n\).

The first result establishes that the true transformation parameter
\(\boldsymbol{\lambda}_0\) is uniquely identified by the population profiled
criterion \(g(\boldsymbol{\lambda})\). This provides the key well-posedness
property for Stage~1: minimizing the profiled criterion targets a unique
population quantity rather than an ambiguous equivalence class. The argument
combines the invertibility of the transformation family with the classical ICA
identifiability result, and the proof is deferred to
Appendix B of Supplemental Material.


\begin{theorem}[Identifiability of the transformation parameter]
\label{thm:identifiability}
Assume the model \eqref{eq:tica_model} and the conditions in
Appendix A of Supplemental Material. Then the profiled population objective
\(g(\boldsymbol{\lambda})\) in \eqref{eq:profile_objective} satisfies
\begin{equation}
g(\boldsymbol{\lambda})\ge 0\ \text{for all }\boldsymbol{\lambda}\in\Lambda,
\qquad
g(\boldsymbol{\lambda}_0)=0,
\qquad
\inf_{\boldsymbol{\lambda}\in\Lambda:\ \boldsymbol{\lambda}\neq \boldsymbol{\lambda}_0}
g(\boldsymbol{\lambda})>0.
\label{eq:g_unique_min}
\end{equation}
Consequently, \(\boldsymbol{\lambda}_0\) is the unique global minimizer of
\(g(\boldsymbol{\lambda})\) over \(\Lambda\).
\end{theorem}

\noindent The detailed proof of Theorem~\ref{thm:identifiability} is given in Appendix B of Supplemental Material. Theorem~\ref{thm:identifiability} implies that Stage~1 targets a unique
population parameter: any misspecified transformation
\(\boldsymbol{\lambda}\neq\boldsymbol{\lambda}_0\) leaves residual dependence
that cannot be removed by linear demixing, and this residual dependence is
captured by the strict positivity in \eqref{eq:g_unique_min}. The role of
invertibility is crucial here: because
\(T_{\boldsymbol{\lambda}}\) is one-to-one on the support, the only way to
achieve mutual information zero after demixing is to undo the correct
componentwise distortion, up to the usual ICA permutation and scaling.

Having established identifiability, we turn to the large-sample behavior of
the Stage~1 estimator \(\widehat{\boldsymbol{\lambda}}_n\) defined in
\eqref{eq:lambda_estimator}. The main technical difficulty is that the empirical
objective \(\widehat g_n(\boldsymbol{\lambda})\) is itself defined through an
inner ICA fit, and thus depends on \(\boldsymbol{\lambda}\) through a
data-dependent demixing matrix. The proof strategy is to show that
\(\widehat g_n\) converges uniformly to \(g\) over \(\Lambda\), that \(g\) is
well-separated around its unique minimizer \(\boldsymbol{\lambda}_0\) as in
\eqref{eq:g_unique_min}, and that \(g\) is locally twice differentiable with
nondegenerate Hessian at \(\boldsymbol{\lambda}_0\). The required uniformity and
smoothness are ensured by the bounded support and regularity conditions in
Appendix A of Supplemental Material, and the detailed proof of Theorem~\ref{thm:stage1} is given in Appendix C of Supplemental Material with supporting lemmas and their proof.

\begin{theorem}[Stage~1 consistency and asymptotic normality]
\label{thm:stage1}
Assume the conditions in Appendix A of Supplemental Material, and let
\(\widehat{\boldsymbol{\lambda}}_n\) be defined by \eqref{eq:lambda_estimator}. Then
\begin{equation}
\widehat{\boldsymbol{\lambda}}_n \xrightarrow{P} \boldsymbol{\lambda}_0.
\label{eq:lambda_consistency}
\end{equation}
If in addition \(g\) is twice continuously differentiable in a neighborhood of
\(\boldsymbol{\lambda}_0\) and the Hessian
\(\mathbf{H}_{\boldsymbol{\lambda}}=\nabla^2 g(\boldsymbol{\lambda}_0) \in \mathbb{R}^{K \times K}\) is positive
definite, then
\begin{equation}
\sqrt{n}\,(\widehat{\boldsymbol{\lambda}}_n-\boldsymbol{\lambda}_0)
\xrightarrow{d}
N\!\left(\mathbf{0},\,
\mathbf{H}_{\boldsymbol{\lambda}}^{-1}\mathbf{V}_{\boldsymbol{\lambda}}\mathbf{H}_{\boldsymbol{\lambda}}^{-1}\right),
\label{eq:lambda_clt}
\end{equation}
where \(\mathbf{V}_{\boldsymbol{\lambda}}\) is the asymptotic covariance of the score
associated with the criterion \(\widehat g_n(\boldsymbol{\lambda})\) evaluated
at \(\boldsymbol{\lambda}_0\).
\end{theorem}
\noindent Theorem~\ref{thm:stage1} justifies optimizing \(\widehat g_n\) in Stage~1:
\eqref{eq:lambda_consistency} shows that the empirical minimizer targets the
unique population minimizer in \eqref{eq:g_unique_min}, while
\eqref{eq:lambda_clt} produces a standard \(\sqrt{n}\)-rate and a quadratic
approximation of the criterion near \(\boldsymbol{\lambda}_0\). Although the
inner ICA fit makes \(\widehat g_n\) non-smooth in finite samples, the
regularity conditions ensure that the resulting map
\(\boldsymbol{\lambda}\mapsto \widehat g_n(\boldsymbol{\lambda})\) behaves as a
regular M-estimation objective in the limit.

Finally, we study Stage~2, in which the demixing matrix is estimated from the
transformed data at \(\widehat{\boldsymbol{\lambda}}_n\). The main question is
whether using \(\widehat{\boldsymbol{\lambda}}_n\) instead of the true
\(\boldsymbol{\lambda}_0\) affects the first-order asymptotics of the demixing
estimator. Since \(\widehat{\boldsymbol{\lambda}}_n\) is \(\sqrt{n}\)-consistent
by \eqref{eq:lambda_clt}, it suffices to establish that the FastICA functional
is sufficiently regular with respect to small perturbations of the underlying
distribution induced by \(\boldsymbol{\lambda}\). This is achieved by combining
continuity properties of \(T_{\boldsymbol{\lambda}}\) (through
\eqref{eq:boxcox_transform} and bounded support) with existing asymptotic theory
for FastICA under whitening.

\begin{theorem}[Stage~2 oracle property]
\label{thm:stage2}
Assume the conditions in Appendix A of Supplemental Material. Let \(\mathbf{W}_0\) denote the
population demixing matrix corresponding to the transformed model at the true
parameter \(\boldsymbol{\lambda}_0\), and let \(\widehat{\mathbf{W}}_n\) be the Stage~2
demixing estimator computed from \(\{T_{\widehat{\boldsymbol{\lambda}}_n}(\mathbf{x}_i)\}_{i=1}^n\).
Then, up to the standard ICA permutation and sign indeterminacies, \(\widehat{\mathbf{W}}_n\) is consistent for \(\mathbf{W}_0\). Moreover, for any fixed row \(\widehat{\mathbf{w}}_{n}\)
of \(\widehat{\mathbf{W}}_n\) corresponding to a pre-specified component,
\begin{equation}
\sqrt{n}\,(\widehat{\mathbf{w}}_{n}-\mathbf{w}_0)\xrightarrow{d}N(\mathbf{0},\boldsymbol{\Sigma}_w),
\label{eq:w_clt}
\end{equation}
where \(\mathbf{w}_0\) is the corresponding row of \(\mathbf{W}_0\) after appropriate permutation
and sign alignment. Importantly, \(\boldsymbol{\Sigma}_w\) coincides with the asymptotic covariance that would be obtained
if the true transformation parameters \(\boldsymbol{\lambda}_0\) were known and ICA were applied directly to the transformed observations
\(\{T_{\boldsymbol{\lambda}_0}(\mathbf{x}_i)\}_{i=1}^n\).
\end{theorem}

A detailed proof of Theorem \ref{thm:stage2}  is provided in Appendix D of Supplemental Material. Theorem~\ref{thm:stage2} establishes an oracle property: replacing the unknown
\(\boldsymbol{\lambda}_0\) by its Stage~1 estimate in \eqref{eq:final_sources}
does not alter the first-order limiting distribution in \eqref{eq:w_clt}. This
formalizes the intuition that Stage~1 learns the transformation sufficiently
fast that Stage~2 effectively operates on data transformed as if by the true
parameter. In particular, inference for demixing directions can proceed using
standard FastICA asymptotic covariance expressions, without additional
variance-inflation terms attributable to the preliminary estimation in
\eqref{eq:lambda_estimator}.

Beyond these pointwise, first--order results, the two–stage procedure also admits
a stronger pathwise guarantee. In particular, under mild strengthening of the
regularity conditions used in Theorems~\ref{thm:stage1} and~\ref{thm:stage2},
we can upgrade convergence in probability of the transformation parameter and
the demixing matrix to joint almost–sure convergence. This yields a global
“eventual correctness’’ result: with probability one, for all sufficiently
large sample sizes, the estimated transformation and demixing directions are
simultaneously arbitrarily close to their population counterparts, up to the usual
ICA permutation and sign indeterminacies.

\begin{theorem}[Joint almost-sure consistency of transformation and demixing]
\label{thm:joint-strong-consistency}
Assume the regularity conditions in Appendix A of Supplemental Material hold,
together with the additional strong-law and continuity conditions for the
Stage~1 criterion and the FastICA functional stated in
Assumptions~\ref{assump:strong-stage1}--\ref{assump:ica-continuity}. Let
\(\boldsymbol{\lambda}_0\) denote the true transformation parameter, and let
\(\mathbf{W}_0\) denote the corresponding population demixing matrix for the transformed
model at \(\boldsymbol{\lambda}_0\). Let \(\widehat{\boldsymbol{\lambda}}_n\) be
the Stage~1 estimator, and let \(\widehat{\mathbf{W}}_n\) be the Stage~2 FastICA demixing
estimator computed from the whitened transformed data
\(\{T_{\widehat{\boldsymbol{\lambda}}_n}(\mathbf{x}_i)\}_{i=1}^n\).

Then, up to the usual ICA permutation and sign indeterminacies,
\[
\widehat{\boldsymbol{\lambda}}_n \xrightarrow{\text{a.s.}} \boldsymbol{\lambda}_0,
\qquad
\widehat{\mathbf{W}}_n \xrightarrow{\text{a.s.}} \mathbf{W}_0,
\]
and hence
\[
\big(\widehat{\boldsymbol{\lambda}}_n,\widehat{\mathbf{W}}_n\big)
\xrightarrow{\text{a.s.}}
\big(\boldsymbol{\lambda}_0,\mathbf{W}_0\big)
\]
jointly, almost surely under the true data-generating process. The detailed
proof, which combines almost-sure uniform convergence of the profiled mutual
information criterion with a strong-law version of the FastICA asymptotic
theory, is given in Appendix E of Supplemental Material.
\end{theorem}

Theorem~\ref{thm:joint-strong-consistency} complements the convergence in
probability and asymptotic normality results by providing a pathwise guarantee
for the entire two–stage procedure. Almost-sure joint convergence of
\(\widehat{\boldsymbol{\lambda}}_n\) and \(\widehat{\mathbf{W}}_n\) is particularly useful
for downstream plug–in constructions (e.g., functionals of the recovered
sources, iterated procedures, or sequential tuning schemes), where stability
along the whole sample path is more natural than statement-by-statement
convergence in distribution. In combination,
Theorems~\ref{thm:identifiability}–\ref{thm:joint-strong-consistency} show that
the proposed AdaptICA framework is well-posed at the population level, admits
efficient \(\sqrt{n}\)-rate estimation for both the transformation and demixing
parameters, and enjoys strong almost-sure consistency guarantees for the
resulting two-step estimators.

\section{Simulation studies}
\label{sec:simulation-studies}

The simulation studies evaluate the finite-sample implications of the theoretical results in Section~\ref{sec:theory}. Examples~1 and~2 are correctly specified under the transformation-based ICA model in \eqref{eq:tica_model} and are used to assess transformation identifiability, Stage~1 estimation, and Stage~2 demixing. Example~3 is a model specification for sensitivity analysis and examines the behavior of AdaptICA when the nonlinear distortions are outside the Box--Cox family. All examples use 300 independent replications with sample size \(n=500\). For AdaptICA, Stage~1 estimates \(\widehat{\boldsymbol{\lambda}}\) by minimizing \(\widehat g_n(\boldsymbol{\lambda})\), and Stage~2 applies FastICA to \(T_{\widehat{\boldsymbol{\lambda}}}(\mathbf x_i)\), producing \(\mathbf u_i^*=\mathbf W^*T_{\widehat{\boldsymbol{\lambda}}}(\mathbf x_i)\) and \(\widehat{\mathbf A}=(\mathbf W^*)^{-1}\). We compare AdaptICA with FastICA, Traditional-BC-FastICA, JADE, and InfoMax. Traditional-BC-FastICA estimates Box--Cox parameters by maximizing marginal Gaussian likelihood, whereas AdaptICA estimates them by minimizing residual dependence after demixing.

Performance is evaluated by Lambda-norm, A-norm, and MI. Lambda-norm, \(\|\widehat{\boldsymbol{\lambda}}-\boldsymbol{\lambda}_0\|\), checks transformation recovery and is used to assess Theorems~\ref{thm:identifiability} and~\ref{thm:stage1}. A-norm, \(\|\widehat{\mathbf A}_{\mathrm{align}}-\mathbf A_0\|_F/\|\mathbf A_0\|_F\), checks mixing-matrix recovery and is used to assess Theorem~\ref{thm:stage2}. MI measures residual dependence among the recovered components. Additional simulation results are provided in Appendix F of Supplemental Material: Subsection F.1 of Supplemental Material first presents figures to demonstrate recovered signal-level to complement the numerical summaries in the following main content, comparing the observed mixtures, AdaptICA recovered components, and true latent sources; Subsection F.2 of Supplemental Material then reports component-wise transformation-parameter summaries for Example~2; and subsection F.3 of Supplemental Material presents an additional sensitivity analysis under misspecified smooth-source nonlinear distortions.

\subsection{Example 1: Scalar transformation parameter}
\label{subsec:sim-example1}

Example~1 considers a scalar Box--Cox transformation shared across all observed variables. The sources are generated as \(S_1\sim t(3)\), \(S_2\sim \mathrm{Uniform}(-1,1)\), and \(S_3\sim \mathrm{Bernoulli}(0.5)-0.5\). The true mixing matrix \(\mathbf A_0\) has rows \((1,2,3)\), \((2,1,1)\), and \((1,3,1)\), and the linear reference data are \(\mathbf X_{\mathrm{linear}}=\mathbf S\mathbf A_0^\top\). In the linear scenario, \(\mathbf X=\mathbf X_{\mathrm{linear}}\), so the target transformation is the identity with \(\lambda_0=1\). In the nonlinear scenario, the observed variables are generated from the inverse Box--Cox transformation with common parameter \(\lambda_0=0.5\), after a column-wise shift to satisfy the Box--Cox domain condition. Thus, Example~1 is a correctly specified scalar-transformation setting.
Table~\ref{tab:lambda_norm_main} shows that AdaptICA has smaller Lambda-norm than Traditional-BC-FastICA in both scenarios of Example~1. This supports Theorems~\ref{thm:identifiability} and~\ref{thm:stage1}: under correct specification, the profiled MI criterion selects a transformation closer to the true parameter than the likelihood-based Box--Cox criterion. The linear scenario also shows that Stage~1 stays close to the identity transformation when no nonlinear correction is needed.

\begin{table}[h!]
\centering
\footnotesize
\caption{Transformation-parameter accuracy for AdaptICA and Traditional-BC-FastICA across the three main simulation examples over 300 replications with \(n=500\). Entries are mean (SD) of Lambda-norm. Smaller values indicate better recovery of the true Box--Cox transformation parameter. An em dash denotes that Lambda-norm is not defined because no true Box--Cox parameter exists in the nonlinear scenario of Example~3. Bold indicates the better value between the two transformation-based methods for each scenario and example.}
\label{tab:lambda_norm_main}
\begin{tabular}{llccc}
\toprule
\textbf{Scenario} & \textbf{Method} & \textbf{Example 1} & \textbf{Example 2} & \textbf{Example 3} \\
\midrule
\multirow{2}{*}{Linear}
& AdaptICA & \textbf{0.197} (0.242) & \textbf{0.016} (0.056) & \textbf{0.005} (0.030) \\
& Traditional-BC-FastICA & 0.546 (0.681) & 0.479 (0.532) & 0.807 (0.424) \\
\midrule
\multirow{2}{*}{Nonlinear}
& AdaptICA & \textbf{0.088} (0.181) & \textbf{0.428} (0.288) & --- \\
& Traditional-BC-FastICA & 0.418 (0.467) & 0.720 (0.755) & --- \\
\bottomrule
\end{tabular}
\end{table}

\begin{table}[h!]
\centering
\footnotesize
\caption{Separation performance across the three main simulation examples over 300 replications with \(n=500\). Lambda-norm is reported separately in Table~\ref{tab:lambda_norm_main}. Entries are mean (SD). Smaller values indicate better performance. Bold indicates the best value in each metric and scenario; when entries are indistinguishable at the reported precision, tied best displayed entries are bolded. Horizontal rules separate the linear and nonlinear scenarios within each example.}
\label{tab:main_simulation_results}
\begin{adjustbox}{max width=\textwidth}
\begin{tabular}{lllcc}
\toprule
\textbf{Example} & \textbf{Scenario} & \textbf{Method} & \textbf{A-norm} & \textbf{MI} \\
\midrule
\multirow{10}{*}{Example 1: Scalar \(\lambda\)}
& \multirow{5}{*}{Linear}
& AdaptICA & 0.068 (0.053) & 0.155 (0.068) \\
& & FastICA & \textbf{0.045} (0.019) & 0.180 (0.077) \\
& & Traditional-BC-FastICA & 0.251 (0.224) & 0.244 (0.056) \\
& & JADE & 0.083 (0.040) & 0.323 (0.105) \\
& & InfoMax & 0.046 (0.020) & \textbf{0.148} (0.065) \\
\cmidrule(lr){2-5}
& \multirow{5}{*}{Nonlinear}
& AdaptICA & \textbf{0.065} (0.083) & \textbf{0.153} (0.064) \\
& & FastICA & 0.207 (0.049) & 0.370 (0.062) \\
& & Traditional-BC-FastICA & 0.257 (0.216) & 0.264 (0.061) \\
& & JADE & 0.261 (0.088) & 0.397 (0.078) \\
& & InfoMax & 0.206 (0.048) & 0.368 (0.061) \\
\midrule
\multirow{10}{*}{Example 2: Vector \(\boldsymbol{\lambda}\)}
& \multirow{5}{*}{Linear}
& AdaptICA & \textbf{0.044} (0.021) & 0.171 (0.076) \\
& & FastICA & \textbf{0.044} (0.021) & 0.175 (0.079) \\
& & Traditional-BC-FastICA & 0.046 (0.022) & 0.237 (0.059) \\
& & JADE & 0.089 (0.038) & 0.326 (0.109) \\
& & InfoMax & 0.044 (0.022) & \textbf{0.143} (0.068) \\
\cmidrule(lr){2-5}
& \multirow{5}{*}{Nonlinear}
& AdaptICA & 0.075 (0.092) & \textbf{0.217} (0.075) \\
& & FastICA & 0.237 (0.149) & 0.332 (0.078) \\
& & Traditional-BC-FastICA & \textbf{0.047} (0.022) & 0.256 (0.068) \\
& & JADE & 0.389 (0.111) & 0.360 (0.100) \\
& & InfoMax & 0.227 (0.138) & 0.342 (0.076) \\
\midrule
\multirow{10}{*}{Example 3: Model specification}
& \multirow{5}{*}{Linear}
& AdaptICA & 0.189 (0.094) & 0.255 (0.023) \\
& & FastICA & 0.186 (0.087) & 0.256 (0.022) \\
& & Traditional-BC-FastICA & 0.338 (0.141) & \textbf{0.251} (0.028) \\
& & JADE & \textbf{0.159} (0.067) & 0.255 (0.022) \\
& & InfoMax & 0.186 (0.091) & 0.259 (0.023) \\
\cmidrule(lr){2-5}
& \multirow{5}{*}{Nonlinear}
& AdaptICA & 0.457 (0.132) & \textbf{0.144} (0.044) \\
& & FastICA & 0.459 (0.135) & 0.200 (0.033) \\
& & Traditional-BC-FastICA & 0.488 (0.130) & 0.219 (0.028) \\
& & JADE & \textbf{0.443} (0.131) & 0.216 (0.035) \\
& & InfoMax & 0.459 (0.134) & 0.204 (0.033) \\
\bottomrule
\end{tabular}
\end{adjustbox}
\end{table}

The separation results in Table~\ref{tab:main_simulation_results} are consistent with Theorem~\ref{thm:stage2}. In the nonlinear scenario, the more accurate transformation estimate for AdaptICA is accompanied by the best A-norm and the best MI. This indicates that estimating \(\boldsymbol{\lambda}_0\) through \(\widehat g_n(\boldsymbol{\lambda})\) leads to transformed data on which the final ICA step can recover a better demixing structure and more independent components.

\subsection{Example 2: Vector transformation parameters}
\label{subsec:sim-example2}

Example~2 uses the same sources, mixing matrix, and sample size as Example~1, but allows a separate Box--Cox transformation parameter for each observed variable. In the linear scenario, the target transformation is \(\boldsymbol{\lambda}_0=(1,1,1)^\top\). In the nonlinear scenario, the true transformation parameter is \(\boldsymbol{\lambda}_0=(0.5,1.0,1.5)^\top\). After forming \(\mathbf X_{\mathrm{linear}}=\mathbf S\mathbf A_0^\top\), each observed variable is transformed using its own inverse Box--Cox parameter. This example checks whether Stage~1 can identify heterogeneous transformation parameters. Additional component-wise summaries for the estimated vector-valued transformation parameter are reported in Section~\ref{app:vector_lambda_details}.

The Lambda-norm results in Table~\ref{tab:lambda_norm_main} favor AdaptICA in both scenarios, extending the empirical support for Theorems~\ref{thm:identifiability} and~\ref{thm:stage1} from a scalar transformation to a vector-valued transformation. This result also reflects the difference between the two transformation criteria: Traditional-BC-FastICA targets marginal Gaussianity, while AdaptICA targets independence after demixing.

Table~\ref{tab:main_simulation_results} shows that AdaptICA gives the smallest MI in the nonlinear scenario. Traditional-BC-FastICA gives the smallest A-norm in this setting, but its Lambda-norm and MI are larger than those of AdaptICA. Thus, the results support the role of \(\widehat g_n(\boldsymbol{\lambda})\) as a profiled dependence criterion for transformation selection. A representative signal-level comparison for Example~2 is shown in  Appendix F.1 of Supplemental Material. The figure illustrates that, after transformation selection and demixing, the recovered components preserve the main qualitative features of the true sources, including continuous heavy-tailed variation and the discrete switching pattern of the Bernoulli-type source.

\subsection{Example 3: Misspecified nonlinear EEG-like setting}
\label{subsec:sim-example3-model-specification}

Example~3 is a model-misspecification sensitivity analysis designed to evaluate AdaptICA when the nonlinear distortions are outside the exact Box--Cox transformation family. It differs from Examples~1 and~2 because the nonlinear scenario is outside the exact Box--Cox transformation family. The source processes use autoregressive coefficients \(\boldsymbol{\rho}=(0.40,0.35,0.30,0.25)\). The innovations are \(\varepsilon_1(t)\sim \mathrm{Uniform}(-\sqrt{3},\sqrt{3})\), \(\varepsilon_2(t)=\{\chi^2_2(t)-2\}/2\), \(\varepsilon_3(t)=\{\chi^2_4(t)-4\}/\sqrt{8}\), and \(\varepsilon_4(t)\sim N(0,1)\). Each source is standardized after generation. The mixing matrix \(\mathbf A_0\in\mathbb R^{4\times4}\) has independent standard normal entries in each replication. In the linear scenario, the target transformation is \(\boldsymbol{\lambda}_0=(1,1,1,1)^\top\). In the nonlinear scenario, the observed variables are distorted by signed power, cubic, and signed logarithmic transformations. Since these distortions are not Box--Cox transformations, the nonlinear scenario does not have a true Box--Cox parameter. A related model-specification sensitivity analysis with misspecified smooth-source nonlinear distortions is reported in Section~\ref{app:misspecified-smooth-simulation}.

The linear scenario in Example~3 checks whether Stage~1 stays close to the identity transformation when the identity is appropriate. Table~\ref{tab:lambda_norm_main} shows that AdaptICA has smaller Lambda-norm than Traditional-BC-FastICA. This supports the stability of the profiled MI criterion, although Example~3 is not part of the correctly specified setting used to verify Theorems~\ref{thm:identifiability} and~\ref{thm:stage1}.

The nonlinear scenario of Example~3 is not a direct check of Theorems~\ref{thm:identifiability}--\ref{thm:joint-strong-consistency}, because the Box--Cox model is misspecified and Lambda-norm is not defined. Table~\ref{tab:main_simulation_results} shows that AdaptICA has A-norm comparable to the competing ICA methods and the smallest MI. Thus, under model misspecification, the transformation step mainly reduces empirical dependence while maintaining similar mixing-matrix recovery. A representative recovered-signal display for this misspecified nonlinear setting is provided in Appendix G.1 of Supplemental Material. Although the fitted Box--Cox family does not coincide with the true nonlinear distortion, the recovered components retain several source-level temporal features, supporting the use of AdaptICA as a dependence-reducing preprocessing and demixing strategy under moderate model misspecification.


\section{Real-Data Analyses}
\label{sec:real_data}

The proposed transformation-based independent component analysis (AdaptICA)
framework is applied to two real neurophysiological signal datasets. The first
analysis considers mu-band power derived from motor-imagery
electroencephalography (EEG) in the BCI Competition IV Dataset~2a. This
application represents a setting in which a scientifically meaningful
nonlinear feature construction changes the measurement scale on which a linear
ICA representation is expected to hold. The second analysis considers
magnetoencephalography (MEG) recordings from the MNE sample-data repository.
For these data, the identity transformation is selected, demonstrating that
AdaptICA retains standard ICA when the original signal scale is already
compatible with linear source separation. Together, the two applications
illustrate the complementary roles of AdaptICA: learning an appropriate
transformation when it improves the independence structure, while preserving
standard ICA when no transformation is supported by the data.

\subsection{Motor-Imagery EEG Analysis}
\label{sec:real_data_BCI}

We first analyze the training session of subject A01 from the BCI Competition
IV Dataset~2a \citep{bcicompiv}, a widely used benchmark for motor-imagery EEG
analysis. The recordings contain 22 scalp EEG channels measured while the
participant imagined movements of the left hand, right hand, feet, and tongue.
These tasks are associated with changes in sensorimotor rhythms over
physiologically relevant cortical regions. Complete information on trial
extraction, signal cleaning, re-referencing, channel grouping, and feature
construction is provided in G.2 of Supplemental Material.

The analysis focuses on mu-band power, corresponding approximately to the
\(8\)--\(13\)~Hz frequency range. Mu rhythms are observed primarily over
sensorimotor cortical regions, and task-related decreases and increases in
mu-band power are commonly interpreted as event-related desynchronization and
synchronization associated with motor preparation and imagined movement
\citep{pfurtscheller1999event}. Mu-band power therefore has direct
physiological relevance and is routinely used as a feature in motor-imagery
brain--computer-interface studies.

\begin{figure}[t]
\centering
\includegraphics[width=\textwidth]{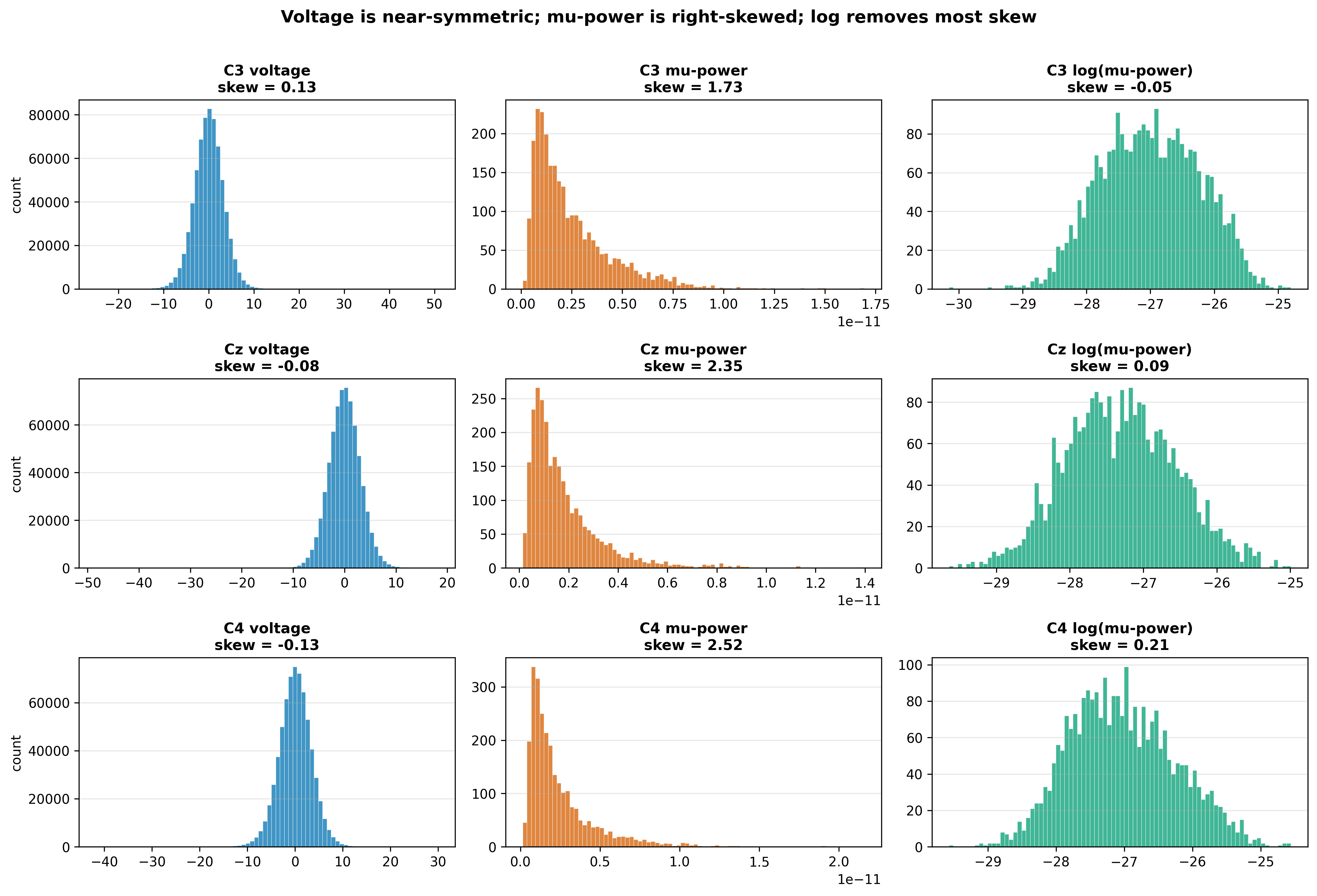}
\caption{Representative marginal distributions of the cleaned EEG signals on
three measurement scales: channel voltage (left), mu-band power obtained by
squaring and locally averaging the \(8\)--\(13\)~Hz band-limited amplitude
(middle), and log mu-band power (right). The comparison illustrates the
nonlinear change in scale induced by the physiologically meaningful power
calculation and motivates estimating an appropriate transformation before
linear source separation.}
\label{fig:voltage_vs_power_main}
\end{figure}

The construction of mu-band power provides the principal motivation for
AdaptICA. Specifically, power is obtained by band-pass filtering the recorded
voltage, squaring the band-limited amplitude, and locally averaging the
resulting signal. This nonlinear componentwise operation is performed after
the latent neural signals have already been mixed at the scalp sensors and
therefore need not preserve the linear source-mixture structure assumed by
standard ICA. Conventional preprocessing typically imposes a prespecified
logarithmic or square-root transformation, often uniformly across all channels
and independently of the subsequent source-separation objective. In contrast,
AdaptICA estimates spatially structured Box--Cox transformation parameters using
the independence criterion, allowing the measurement scale of the power
representation to be adjusted before linear demixing.
Figure~\ref{fig:voltage_vs_power_main} illustrates the substantial change in
measurement scale induced by the physiologically meaningful power calculation
and thereby motivates transformation-based source separation.

After preprocessing and temporal subsampling, the resulting dataset contains
\(n=13{,}451\) observations measured over \(T=22\) channels. We considered
several candidate latent dimensions together with a nested family of
anatomically structured channel groupings. The proposed model-selection
procedure selected \(\widehat r=10\) and \(\widehat K=7\). Details of the
candidate models, tuning procedure, and numerical results are provided in
Supplemental Appendix G.3 of Supplemental Material.

The two selected quantities have distinct interpretations. The value
\(\widehat r=10\) specifies the dimension of the final latent representation,
so that ten demixed component directions are retained from the original
22-channel recording. This value should not be interpreted as implying that
the brain contains exactly ten physiological generators. Rather, it provides a
parsimonious representation of the dominant independent structure identified
in the analyzed recording.

The selected value \(\widehat K=7\) partitions the 22 EEG channels into seven
spatial groups, with channels in the same group sharing one Box--Cox
transformation parameter. The estimated transformation is $\widehat{\boldsymbol{\lambda}}_{\mathrm{EEG}}
=
(0.115,\,-0.509,\,-0.771,\,-1.000,\,0.258,\,0.873,\,0.055)^\top,$
where the seven entries correspond to the anterior-left, anterior-right,
central-left, central-right, midline, posterior-left, and posterior-right
channel groups, respectively. Thus, \(\widehat r=10\) determines the dimension
of the recovered latent representation, whereas \(\widehat K=7\) determines
the spatial resolution of the transformation. Relative to the identity value
\(\lambda=1\), values below one increasingly compress large power observations,
with values near zero corresponding approximately to a logarithmic
transformation. The substantial variation among the seven estimates indicates
that the adjustment required for mu-band power differs across scalp regions,
supporting a spatially adaptive transformation rather than a single
transformation applied uniformly to all channels.

For the selected specification
\((\widehat r,\widehat K)=(10,7)\), AdaptICA reduced the mutual information among
the recovered components from \(0.07668\) to \(0.04696\), corresponding to a
\(38.8\%\) reduction. Distance correlation, which was not used to estimate the
transformation parameters, decreased from \(0.15064\) to \(0.08511\), a
\(43.5\%\) reduction. The consistent improvements under both dependence
measures indicate that the fitted transformation produces a more nearly
independent latent representation and that the improvement is not specific to
the mutual-information criterion used for estimation.

These improvements in statistical independence among the recovered components
are accompanied by greater spatial focality and task specificity under AdaptICA
than under standard ICA in the physiologically targeted hand-imagery
comparisons. To evaluate whether the
reduction in residual dependence is accompanied by clearer sensorimotor
organization, we conducted a physiologically motivated comparison based on the
expected spatial patterns of motor imagery. Right-hand imagery was evaluated
over the contralateral C3 region, left-hand imagery over the contralateral C4
region, and feet imagery near the central midline around Cz. For each method
and target region, we selected the component with the greatest target-region
loading concentration and examined both its scalp topography and its
cue-aligned, baseline-normalized trajectory.

\begin{figure}[h!]
\centering
\includegraphics[width=\textwidth]{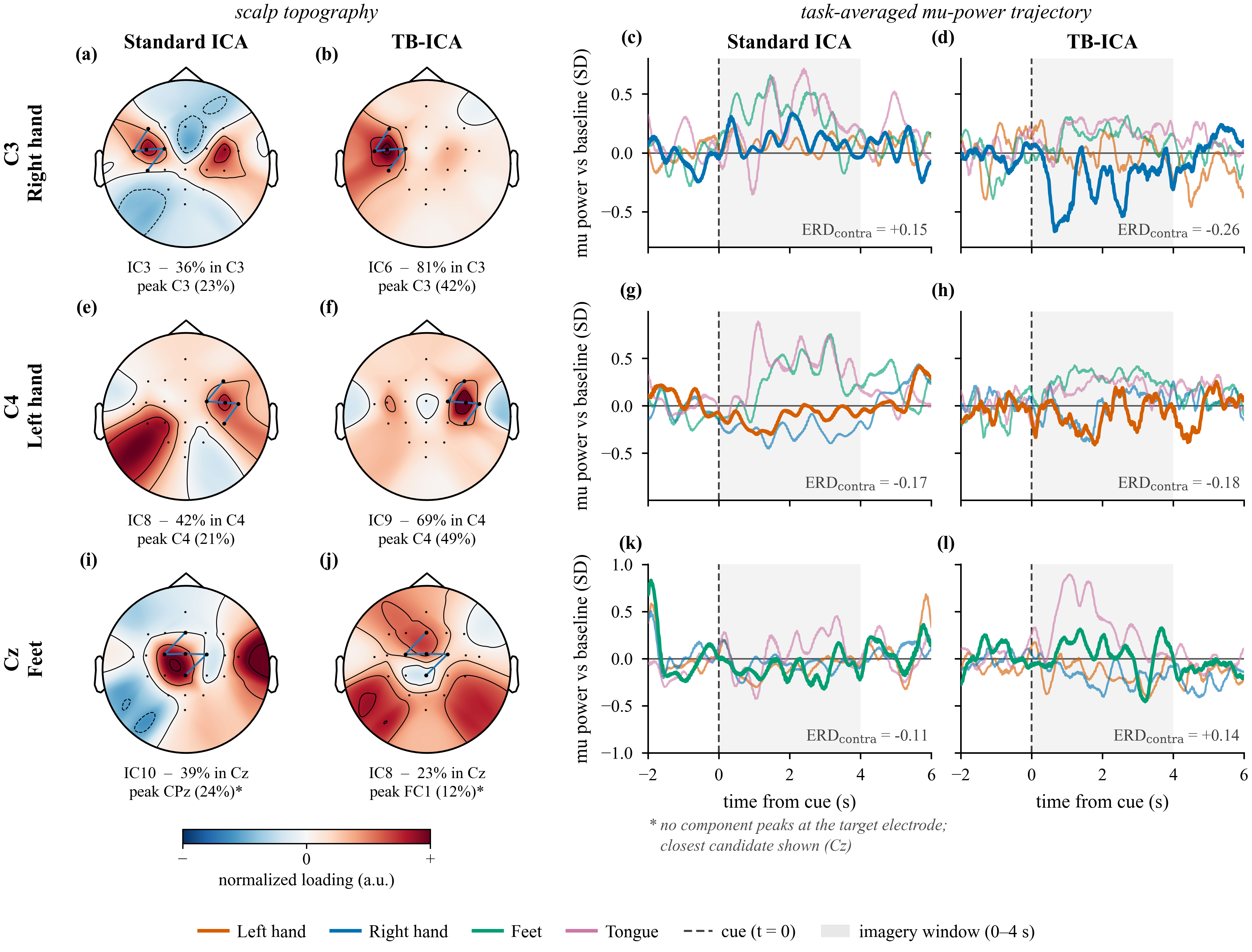}
\caption{Recover component comparisons of standard ICA and AdaptICA for
right-hand, left-hand, and feet motor imagery. Each row compares the component
selected separately within each method using a physiologically motivated
target-region criterion: C3 for right-hand imagery, C4 for left-hand imagery,
and Cz for feet imagery. The first two columns show normalized scalp-loading
maps, and the last two columns show task-averaged component trajectories
relative to the pre-cue baseline. The target-task trajectory is emphasized by
a thicker line. The dashed vertical line marks cue onset, and the shaded region
denotes the \(0\)--\(4\)-second imagery interval. Component signs are oriented
consistently with the target-region loading so that a negative target-task
deflection can be interpreted as an ERD-like decrease.}
\label{fig:eeg_motor_main}
\end{figure}

As shown in Figure~\ref{fig:eeg_motor_main}, AdaptICA produced a clearly
C3-centered component whose right-hand trajectory decreased during the imagery
interval, with an imagery-window summary of \(-0.26\). By comparison, the
standard-ICA candidate did not exhibit a corresponding ERD-like decrease, with
an imagery-window summary of \(+0.15\). For left-hand imagery, both methods
produced similar imagery-window decreases, with values of \(-0.17\) for
standard ICA and \(-0.18\) for AdaptICA, but the AdaptICA component was more
strongly localized around the contralateral C4 region. In contrast, the
feet-related comparison did not favor AdaptICA. The physiological gains should
therefore be interpreted as component- and task-specific rather than uniform.
Collectively, the results suggest that transformation learning can reveal
selected sensorimotor patterns that are partially obscured when standard ICA is
applied directly to the nonlinear mu-band-power scale, without implying that
every recovered component must become more physiologically informative.

\begin{figure}[h!]
\centering
\includegraphics[width=\textwidth]{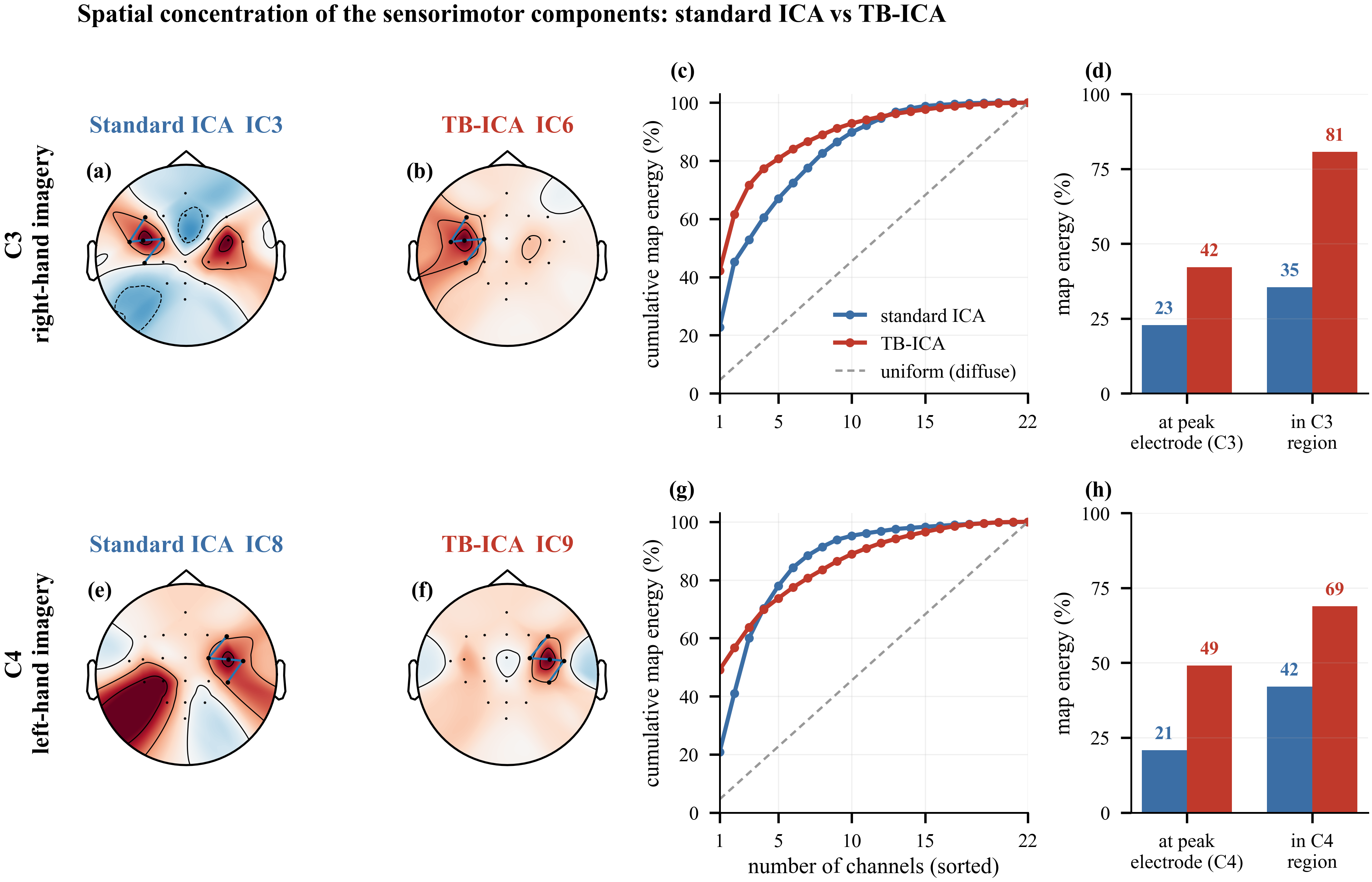}
\caption{Quantitative comparison of spatial concentration for the components
selected for right-hand imagery over C3 and left-hand imagery over C4. Panels
(a), (b), (e), and (f) show the normalized scalp-loading maps. Panels (c) and
(g) show cumulative map energy after the channels are sorted from largest to
smallest squared loading; the diagonal reference represents uniformly diffuse
energy across the 22 channels. Panels (d) and (h) compare the percentage of
total map energy at the target electrode and within the prespecified target
region. AdaptICA yields greater target-electrode and target-region concentration
for both hand-imagery comparisons, indicating more spatially focal
contralateral sensorimotor representations.}
\label{fig:eeg_focal_main}
\end{figure}

The greater spatial focality of the hand-related AdaptICA components is
quantified in Figure~\ref{fig:eeg_focal_main}. For right-hand imagery, the
proportion of total map energy at the peak C3 electrode increased from \(23\%\)
under standard ICA to \(42\%\) under AdaptICA, while the proportion contained in
the prespecified C3 region increased from approximately \(35\%\) to \(81\%\).
For left-hand imagery, the corresponding concentration at the peak C4
electrode increased from \(21\%\) to \(49\%\), and the concentration within the
C4 region increased from \(42\%\) to \(69\%\). The cumulative map-energy curves
further show how rapidly the component loading energy accumulates over the
channels with the largest absolute loadings.

These differences clarify the practical effect of applying standard ICA
directly to the nonlinear mu-band-power scale. Because standard ICA must
represent the derived power variables using a purely linear demixing model,
localized motor-related variation may be distributed over a broader scalp
pattern or combined with diffuse scale-induced variation. By adapting the
measurement scale across scalp regions before demixing, AdaptICA can isolate
selected contralateral hand-related patterns more sharply. The additional
physiological interpretation is therefore not that AdaptICA uniformly produces
stronger activation, but that it can yield a more spatially focused and
task-specific representation of particular sensorimotor processes.

Taken together, the EEG results demonstrate that AdaptICA provides more than a
reduction in global dependence among the recovered components. By accounting
for regional heterogeneity in the nonlinear mu-band-power scale, the method can
produce selected latent components with clearer spatial localization and
task-related interpretation. At the same time, the feet-related result confirms
that such gains are not expected to occur uniformly across all tasks or
components. A complete component-by-region and component-by-task assessment,
which places these targeted examples within the full ten-component
decomposition, is provided in Supplemental
Figure~\ref{fig:eeg_matrices_app}. The following MEG analysis provides a
complementary case in which no transformation is selected and the standard ICA
representation is retained.

\subsection{MEG Analysis}
\label{sec:real_data_MEG_summary}

We next apply the model-selection framework to continuous MEG recordings from
the MNE sample-data repository
\citep{gramfort2013meg,mnesampledata}. These recordings were collected during
an auditory--visual sensorimotor experiment and provide a complementary
application because MEG sensors measure magnetic fields generated by neural
currents. Such fields are commonly represented through approximately linear
electromagnetic superposition, making standard linear ICA a scientifically
meaningful candidate for this signal representation. Detailed preprocessing
and numerical results are provided in Supplemental
Appendix~\ref{app:meg_analysis}.

The candidate model class explicitly included $\lambda=1$,
which represents the identity transformation and is therefore equivalent to
standard ICA after centering. The model-selection procedure selected this
identity model. Standard FastICA attained mutual information \(0.013928\),
compared with \(0.014780\) for the optimized nonidentity transformed candidate
and \(0.028055\) for a fixed logarithmic transformation. Thus, the optimized
nonidentity transformation increased residual dependence by approximately
\(6.1\%\), whereas the uniform logarithmic transformation more than doubled
the residual mutual information. The selection of \(\lambda=1\) has a meaningful scientific interpretation. It
indicates that, for the analyzed MEG recordings, the original measurement
scale is more compatible with a linear independent-component decomposition
than the alternative transformed scales. This empirical result is consistent
with the linear-superposition interpretation of MEG measurements. It also
shows why transformations should not be imposed automatically: unnecessary
transformation can obscure rather than improve the latent source structure.

The EEG and MEG analyses therefore provide complementary demonstrations of the general contribution of AdaptICA. For EEG mu-band power, the scientifically
motivated square operation changes the observed measurement scale, and region-specific Box--Cox transformations yield components with substantially
less residual dependence. For the MEG data, the selected value
\(\lambda=1\) retains standard ICA because the original scale is empirically
more appropriate. AdaptICA is consequently more than an additional preprocessing method. It is a
flexible, unified model-selection framework that contains standard ICA as a special case, determines whether a transformation is supported by the data,
and estimates the required transformation when the identity model is
inadequate. More generally, the method extends the applicability of
interpretable linear ICA to signal representations affected by unknown
componentwise distortions without sacrificing the validity or practical
advantages of standard ICA in settings where the linear model already fits
well.

\section{Conclusion and Future directions}
This paper proposed a transformation-based ICA framework that incorporates componentwise Box--Cox transformations into a standard linear ICA pipeline. By selecting transformation parameters through a profiled mutual-information criterion, TB-ICA directly targets statistical independence while retaining the structure and practicality of classical ICA algorithms. The transformation step provides a data-driven correction for moderate nonlinearities and post-nonlinear mixing, making TB-ICA a natural enhancement for EEG, MEG, fMRI, and related source-separation workflows. Theoretically, we established identifiability of the transformation parameter, consistency and asymptotic normality of the Stage~1 estimator, an oracle property for the Stage~2 FastICA demixing directions, and joint almost-sure consistency of the transformation parameters and demixing matrix.

Several future directions remain. First, richer transformation families, such as spline-based or flow-based monotone maps, could improve flexibility while raising new questions about identifiability, regularization, and interpretability. Second, extending TB-ICA to temporally dependent signals and multi-subject studies may improve robustness and cross-subject comparability. The identified independent components could also serve as low-dimensional and interpretable latent predictors in downstream dynamic regression models, including varying-coefficient frailty and mixed-effects formulations \citep{hung2022varying,jalili2025scalable}. Third, the relationship of TB-ICA to identifiability frameworks based on temporal dependence or auxiliary variables \citep{hyvarinen2016auxiliary,hyvarinen2017timecontrastive,lyu2022finite} should be further clarified. Finally, replacing mutual information with more robust or scalable dependence measures \citep{chen2008order,ilmonen2011semiparametric,reyhani2012consistency,chen2006efficient,matteson2017dCovICA} and developing online or adaptive variants would broaden TB-ICA's applicability to real-time neurophysiology, large-scale multichannel systems, and other domains involving nonlinear marginal distortions.



{\setstretch{1}
\bibliographystyle{plainnat}
\bibliography{references}
}

\end{document}